\documentclass{article}

\usepackage[dvips]{graphicx}
\newcommand{\ve}[1]{\mathbf{#1}}
\newcommand{\Aen}{{\cal A}}
\newcommand{\Ben}{{\cal B}}
\newcommand{\Cen}{{\cal C}}
\newcommand{\Den}{{\cal D}}
\newcommand{\Een}{{\cal E}}
\newcommand{\Gen}{{\cal G}}

\newtheorem{example}{Example}
\newtheorem{definition}{Definition}
\newtheorem{lemma}{Lemma}
\newtheorem{corollary}{Corollary}
\newtheorem{remark}{Remark}
\newtheorem{theorem}{Theorem}
\newcommand{\defeq}{\stackrel{\triangle}{=}}
\newcommand{\qed}{\fbox{}}

\begin{document}
\title{Average Coset Weight Distribution 
    of Combined LDPC Matrix Ensemble}
\author{Tadashi Wadayama
\thanks{T.Wadayama is with 
  Department of Computer Science, 
  Nagoya Institute of Technology, Nagoya, 466-8555, Japan.
  (e-mail:wadayama@nitech.ac.jp).
  The work will be presented in part at 
  Hawaii, IEICE and SITA Joint Conference on Information Theory, May, 2005.
 } }
 \maketitle

\begin{abstract}
In this paper,
the average coset weight distribution (ACWD) of structured 
ensembles of LDPC (Low-density Parity-Check) matrix, 
which is called {\em combined ensembles}, is discussed.
A combined ensemble is composed of a set of simpler ensembles
such as a regular bipartite ensemble.
Two classes of combined ensembles have prime importance;
a {\em stacked ensemble} and a {\em concatenated ensemble}, which 
consists of set of stacked matrices and  concatenated matrices, respectively.
The ACWD formulas of these ensembles is shown in this paper.
Such formulas are key tools to evaluate the ACWD of a complex combined ensemble.
From the ACWD of an ensemble,
we can obtain some detailed properties of a code (e.g., weight of coset leaders) which
is not available from an average weight distribution.
Moreover, it is shown that the analysis based on the ACWD 
is indispensable to evaluate the average weight distribution
of some classes of combined ensembles.
\end{abstract}

\section{Introduction}
Ensemble analysis on weight distributions of low-density parity-check (LDPC) codes
gives us insights on  properties  of an instance of the ensemble.
For example, by using the ML-performance bound presented by Miller and Burshtein \cite{Miller}, 
we can derive an ML-performance bound from the average weight distribution of
an ensemble. The typical minimum distance \cite{Gal63} can be also derived from an
average weight distribution.

The first work on the average weight distribution of a regular  LDPC code ensemble
has been made by Gallager \cite{Gal63}. 
Recently, Litsyn and Shevelev \cite{LS02}, \cite{LS03} presented
the average weight distribution of  several regular/irregular LDPC code ensembles
and their asymptotic behaviors.
Burshtein and Miller \cite{MV04} showed methods for asymptotic enumeration and
have derived the average weight distribution of some irregular LDPC ensembles
using their technique. Di, Richardson, and  Urbanke \cite{dru} showed that
the saddle point method is useful to derive the asymptotic growth rate of
an average weight distribution.

The average coset weight distribution (ACWD) of the Gallager ensemble has been 
introduced by the author \cite{wadayama0} \cite{wadayama1}
and the ACWD formulas of the Gallager ensemble has been derived.
These papers show that the ACWD of an ensemble indicates
some detailed properties of a code (e.g., weight of coset leaders) which
is not available from an average weight distribution.

Structured LDPC codes are known to be a promising class of codes which
attains excellent trade-offs between decoding complexity versus 
decoding performance.
An instance of a structured LDPC code can be regarded as
a regular or an irregular LDPC code with additional constraints on edge connections.
This class of LDPC codes includes some well known codes 
such as repeat accumulate codes (RA code) \cite{Divsalar},
irregular repeat and accumulate codes (IRA code) \cite{Jin},
and multi-edge type LDPC codes \cite{multiedge}, etc..

For design and analysis of structured LDPC codes,  
knowledge on the average weight distribution
and the ACWD is desirable.  For example, the typical minimum distance may be used
to optimize the structure of a code.
In this paper, we discuss ACWDs of
some classes of structured LDPC codes.
Precisely speaking, our main focus is on ensembles of combined matrices, which is called
{\em combined ensemble}. A party check matrix of a structured LDPC codes can be
considered as a combined matrix.
A combined ensemble is composed of a set of simpler ensembles
such as a regular bipartite ensemble. 
Two classes of combined ensemble have prime importance;
a {\em stacked ensemble} and a {\em concatenated ensemble}, which 
consists of a set of stacked matrices and  concatenated matrices, respectively.
The ACWD formulas of these ensembles will be derived in this paper.
Such formulas are key tools to evaluate the ACWD of a complex combined ensemble.

The organization of the paper is as follows.
Section \ref{sec-pre} gives basic notations used throughout the paper and 
presents some known results on the ACWD of several ensembles.
Section \ref{sec-sym} shows symmetric properties of an ensemble which are required 
for the analysis of a combined ensemble. Section \ref{sec_combined} presents
the ACWD formulas for combined ensembles. 
In Section \ref{sec_asymptotic},
asymptotic behaviors of the ACWD and the typical coset weight 
are discussed.

\section{Preliminaries}
\label{sec-pre}
In this section, we will briefly review the definition of the
ACWD according to \cite{wadayama0} \cite{wadayama1} and some known results.

\subsection{Average coset weight distributions}

Let $n$ and $m$ be positive integers, which are called the {\em column size} 
and the {\em row size}, respectively.
For a given binary $m \times n$ parity check matrix $H$ and 
a vector $\ve s \in F_2^m$ ($F_2$ is the Galois field with two elements),
the set $C(H,\ve s)$ is defined by $\{\ve x \in F_2^n: H \ve x^t = \ve s \}$,
which is called the {\em coset} corresponding 
to $\ve s$ ($\ve s$ is called a {\em syndrome}).
The definition of the coset is slightly different from the traditional one
because the rank of $H$ is not necessarily $m$.
The number of vectors of weight $w$ in the set $C(H,\ve s)$
is denoted by $A_w (H, \ve s)$:
\begin{equation}
  A_w (H, \ve s) \defeq \#\{\ve x \in Z^{(n,w)}: H \ve x^t = \ve s \},\quad w \in [0,n],
\end{equation}
where $Z^{(n,w)}$ denotes the set $\{\ve z \in F_2^n: |\ve z| = w \}$.
In this paper,
the size of a set $A$ is represented by $\#A$ and the Hamming weight of a vector 
$\ve x$ is denoted by $|\ve x|$.
The notation $[a,b]$ denotes the set of consecutive integers
from $a$ to $b$.
The set of numbers $\{A_w (H, \ve s)\}_{w=0}^n$ is referred to as 
the {\em coset weight distribution} of $C(H,\ve s)$.

We then consider the average of $A_w (H, \ve s)$ over an ensemble of 
parity check matrices.
For a given ensemble $\Gen$ of parity check matrices,
the ACWD $\{\tilde A_w(\ve s)\}_{w=0}^n$ is given by
\begin{eqnarray} \nonumber
  \tilde A_w(\ve s) &\defeq& \sum_{H \in \Gen} P(H) A_w (H, \ve s) \\ \nonumber
  &=& \sum_{H \in \Gen} P(H) \#\{\ve x \in Z^{(n,w)}: H \ve x^t = \ve s\} \\
  &=& \sum_{H \in \Gen} P(H) \sum_{\ve z \in Z^{(n,w)}} I[H \ve z^t = \ve s],
\end{eqnarray}
where 
$I[condition]$ gives 1 if $condition$ is true; otherwise it gives 0.
Throughout the paper, we consider the ensembles such that
every matrix in the ensemble is associated with uniform probability 
$P(H) = 1/(\# \Gen)$.
By changing the order of the summations, 
we can simplify the above equation as follows:
\begin{eqnarray} \nonumber
  \tilde A_w(\ve s) 
  &=& \sum_{\ve z \in Z^{(n,w)}} \sum_{H \in \Gen} P(H)  I[H \ve z^t = \ve s] \\
  \label{acwd}
  &=& \sum_{\ve z \in Z^{(n,w)}} \frac{\#\{H\in \Gen: H \ve z^t = \ve s \}}{\# \Gen}.
\end{eqnarray}

This equation means that 
evaluation of $\tilde A_w(\ve s)$ is equivalent to 
enumeration of the number of parity check matrices in the set
$\{H\in \Gen: H \ve z^t = \ve s \}$.
In this paper,  
when we need to indicate the dependency on
an ensemble explicitly,
the ACWD of an ensemble $\Gen$ is denoted by $\tilde A_w^{(\Gen)}(\ve s)$.
% The number of columns (resp. rows) of a matrix in an ensemble $\Een$
% is said to be the column (row) size of $\Een$.

\subsection{Known results on ACWD formulas}
\label{knownacwd}
The ACWD formulas of several ensembles of LDPC matrices have been already derived
\cite{wadayama0} \cite{wadayama1} \cite{dru} \cite{shimo04}.
We here review some known results.

The Gallager ensemble \cite{Gal63} consists of $(j,k)$-regular binary matrices
where $j$ and $k$ denotes the column weight and the row weight of a matrix, respectively.
The ACWD of the Gallager ensemble has been proved 
in \cite{wadayama0} \cite{wadayama1}.

\begin{lemma}[ACWD of Gallager ensemble\cite{wadayama0}\cite{wadayama1}]
  \label{averagethorem}
  Let $\Gen$ be a Gallager ensemble such that any matrix in $\Gen$ is 
  $m \times n$-binary matrix and $(j,k)$-regular.
  The ACWD of $\Gen$ is given by
  \begin{equation} \label{gallagerACWD}
    \tilde A_w(\ve s) = {n \choose w} \prod_{i=1}^j \frac{1}{{n \choose w}} 
    \left[
      \alpha_k^{(m/j)-|\ve s_i|}(x) \beta_k^{|\ve s_i|}(x)
      \right]_{w}
  \end{equation}
  for $w \in [0,n], \ve s \in F_2^m$.
  The vector $\ve s_i \in F_2^{m/j} (i \in [1,j])$ is sub-syndrome which satisfies
  $\ve s = (\ve s_1, \ve s_2,\ldots, \ve s_j)$.
The notation $[f(x)]_w$ denotes the coefficient of a polynomial $f(x)$ corresponding
to $x^w$.
The polynomials $\alpha_k(x)$ and $\beta_k(x)$ are defined by
\begin{eqnarray}
  \alpha_k(x) &=& \frac{(1+x)^k+(1-x)^k}{2} \\ \label{oddenumerator}
  \beta_k(x) &=& \frac{(1+x)^k-(1-x)^k}{2},
\end{eqnarray}
which are the weight enumerators of the even and the odd weight codes of length
$k$, respectively. 
\hfill\qed
\end{lemma}

The main difference between the average weight distribution
derived by Gallager \cite{Gal63} and (\ref{gallagerACWD}) is the use of 
the odd weight enumerator (\ref{oddenumerator}).

The next lemma gives the ACWD of a constant row weight ensemble, which is 
referred to as Ensemble E in \cite{LS02}. The ensemble consists of all binary 
matrices whose row weight is exactly $k$.
\begin{lemma}
[ACWD of constant row weight ensemble\cite{wadayama1}\cite{wadayama2}]
Let $\Gen$ be a constant row weight ensemble such that any matrix in $\Gen$ is 
$m \times n$-binary matrix and has constant row weight $k$.
The ACWD of $\Gen$ is given by
\begin{equation}
\tilde A_w(\ve s) =
  \left[
    \gamma(n,k,w)
  \right]
  ^{m-|\ve s|}
  \left[
    1 - \gamma(n,k,w)
  \right]
  ^{|\ve s|} {n \choose w}
\end{equation}
for $w \in [0,n], \ve s \in F_2^m$, where $\gamma(n,k,w)$ is defined by
\begin{equation}
\gamma(n,k,w) \defeq \frac{1}{{n \choose k}}
    \sum_{i=0}^{\lfloor k/2 \rfloor} {w \choose 2 i}{n-w \choose k - 2i}.
\end{equation}
\hfill\qed
\end{lemma}

The ACWD of a regular bipartite graph ensemble (abbreviated as a {\em bipartite ensemble}) 
has been implicitly used to 
prove average weight distribution of IRA code ensembles in \cite{dru} and
tripartite graph ensembles in \cite{shimo04} \cite{Ikegaya}. 
Since these papers do not contain the explicit proof of the
next lemma, the next lemma is presented with the proof, which
is given in Appendix.
\begin{lemma}
[ACWD of regular bipartite ensemble\cite{dru}\cite{shimo04}]
\label{bipartite}
Let $\Gen$ be a regular bipartite  ensemble such that any matrix in $\Gen$ is 
$m \times n$-binary matrix and corresponds to a $(j,k)$-regular bipartite graph
where $j$ and $k$ denote the variable node degree and the check node degree, 
respectively.
Note that the equality $j n = k m$ must be satisfied.
The ACWD of $\Gen$ is given by
\begin{equation}
  \tilde A_w(\ve s) =   \frac{\left[
        \alpha_k^{m-|\ve s|}(x) \beta_k^{|\ve s|}(x)
      \right]_{w j}}{{n j \choose w j}} {n \choose w}
\end{equation}
for $w \in [0,n], \ve s \in F_2^m$. \\
(Proof) See Appendix. \hfill\qed
\end{lemma}

\section{Symmetric Properties of Ensembles}
\label{sec-sym}
In order to deal with combined ensembles introduced in Section \ref{sec_combined}, 
we need to define some symmetric properties of an ensemble.
In this section, two symmetric properties of an ensemble will be
defined. Then, we will discuss how to extend a non-symmetric
ensemble to a symmetric one.

\subsection{Column and row symmetric ensembles}
The definition of the column symmetric property is given as follows.
\begin{definition}[Column symmetric property]
\label{defcolumnsym}
If the equality 
\begin{equation} \label{columnsymmetric}
\# \{H \in \Gen: H \ve z_1^t = \ve s \}
= \# \{H' \in \Gen: H' \ve z_2^t = \ve s \}
\end{equation} 
holds for any $\ve s \in F_2^m$, $w \in [0,n]$,
$\ve z_1 \in Z^{(n,w)}$, and $\ve z_2 \in Z^{(n,w)}$,
then the ensemble $\Gen$ is {\em column symmetric}.
\hfill\qed
\end{definition}
The above definition means that 
$\# \{H \in \Gen: H \ve z^t = \ve s \}$ depends only on the 
weight of $\ve z$ if an ensemble $\Gen$ is column symmetric.
From the symmetric property of the ensemble (\ref{columnsymmetric}),
the equation (\ref{acwd}) can be rewritten into a simpler form:
\begin{eqnarray} \nonumber
  \tilde A_w(\ve s) 
  &=&\sum_{\ve z \in Z^{(n,w)}} \frac{\#\{H\in \Gen: H \ve z^t = \ve s \}}{\# \Gen} \\
  &=& \frac{\#\{H\in \Gen: H \ve z^t = \ve s \}}{\# \Gen} {n \choose w}
\end{eqnarray}
if $\Gen$ is column symmetric. 
For example, all ensembles discussed in Section \ref{knownacwd} are column symmetric.

An important example of an ensemble which is not column symmetric is 
a {\em single matrix ensemble}. Let $\Een(H)$ be an ensemble which includes 
$m \times n$ binary matrix $H$ as an only member. Thus, the probability
$P(H) = 1$ is assigned to $H$.
Such an ensemble $\Een(H)$ is called a single matrix ensemble.
From this definition, it is evident that the equation 
$
  \tilde A_w(\ve s)  = \tilde A_w(H,\ve s)
$
holds for a single matrix ensemble $\Een(H)$.
In general, $\Een(H)$ is not column symmetric.

The row symmetric property, which is defined below, 
greatly simplifies the ensemble analysis presented in 
Section \ref{sec_combined}.
\begin{definition}[Row symmetric property]
If 
the equality 
\begin{equation}
\# \{H \in \Gen: H \ve z = \ve s_1 \}
= \# \{H' \in \Gen: H' \ve z = \ve s_2 \}
\end{equation} 
holds
for any $\ve z \in F_2^n$, $\sigma \in [0,m]$,
$\ve s_1 \in Z^{(m,\sigma)}$, and $\ve s_2 \in Z^{(m,\sigma)}$,
then the ensemble $\Gen$ is {\em row symmetric}.
\hfill\qed
\end{definition}
From this definition, it is evident that
the equation
\begin{eqnarray}
  \tilde A_w(\ve s_1) \nonumber
  &=& \sum_{\ve z \in Z^{(n,w)}} \frac{\#\{H\in \Gen: H \ve z^t = \ve s_1 \}}{\# \Gen} \\ \nonumber
  &=& \sum_{\ve z \in Z^{(n,w)}} \frac{\#\{H\in \Gen: H \ve z^t = \ve s_2 \}}{\# \Gen} \\
  &=& \tilde A_w(\ve s_2) 
\end{eqnarray}
holds if $|\ve s_1| = |\ve s_2|$.
This equation implies that 
the ACWD $\tilde A_w(\ve s)$ depends only  on the 
weight of $\ve s$ if an ensemble $\Gen$ is row symmetric.

For a row symmetric ensemble, we here introduce 
$\tilde B_w(\sigma) (w \in [0,n], \sigma \in [0,m])$ which is defined by
\begin{equation}
  \tilde B_w(|\ve s|) \defeq \tilde A_w(\ve s).  
\end{equation}
The distribution $\tilde B_w(\sigma)$ is also called the ACWD of the ensemble.
One of the advantages of a row symmetric ensemble is that
we only need to handle $(n+1) (m+1)$-numbers of $\tilde B_w(\sigma) 
(\sigma \in [0,m], w \in [0,n])$ instead of 
$(n+1) 2^m$-numbers of $\tilde A_w(\ve s)(\ve s \in F_2^m, w \in [0,n])$ 
when we deal with the ACWD of a row symmetric ensemble.
%Therefore, for simplicity, we use $\tilde B_w(\sigma)$ instead of $\tilde A_w(\ve s)$
%when we deal with a row symmetric ensemble.

\begin{example}
  The bipartite ensemble is row symmetric. Thus, the ACWD of $(j,k)$-regular 
  bipartite ensemble can be expressed by
\begin{equation} \label{acwdbipartite}
  \tilde B_w(\sigma) =   \frac{\left[
        \alpha_k^{m-\sigma}(x) \beta_k^{\sigma}(x)
      \right]_{w j}}{{n j \choose w j}} {n \choose w}.
\end{equation}
\hfill\qed
\end{example}

\subsection{Column shuffled ensemble}

Assume that $\Gen$ (column size $n$ and row size $m$)
is an ensemble which is not column symmetric.
The {\em column shuffled ensemble} $\Gamma_c(\Gen)$ is an
enlarged ensemble based on $\Gen$, which is defined by
\begin{equation}
  \Gamma_c(\Gen) \defeq  \{\rho(H): H \in \Gen,\rho \in \pi_n \},
\end{equation} 
where $\pi_n$ is the set of all column permutations for $n$ columns.
Thus, $\#\Gamma_c(\Gen) = n! \#\Gen$. We here regard $\Gamma_c$ as 
a column shuffling operator
which acts on an ensemble. 
The next lemma shows that the column shuffling yields
a column symmetric ensemble from any ensemble.
\begin{lemma}[Symmetric property of column shuffled ensemble]
  $\Gamma_c(\Gen)$ is column symmetric. \\
(Proof)
For any $\ve z_1 \in F_2^n$ and $\rho \in \pi_n$,
let $\ve z_2 = \rho(\ve z_1)$. For any $\ve s \in F_2^m$,
we have the following equality:
\begin{eqnarray} \nonumber
  \#\{H \in \Gamma_c(\Gen) \hspace{-1mm}:\hspace{-1mm} H \ve z_2 = \ve s\}
  \hspace{-3mm}
  &=& \hspace{-3mm} 
  \#\{H \in \Gamma_c(\Gen)\hspace{-1mm}:\hspace{-1mm} H \ve \rho(\ve z_1) = \ve s\} \hspace{5mm}\\ \nonumber
  &=& \hspace{-3mm} \#\{H \in \Gamma_c(\Gen)\hspace{-1mm}:\hspace{-1mm} \rho(H) \ve z_1 = \ve s\}.
\end{eqnarray}
The second equality follows from 
the equality $H \ve \rho(\ve z_1)=\rho(H) \ve z_1$.
There is one to one correspondence between
$\rho(H)$ and $H$. This fact leads to
\begin{equation}
  \#\{H \in \Gamma_c(\Gen): H \ve z_2 = \ve s\}
  = \#\{H' \in \Gamma_c(\Gen): H' \ve z_1 = \ve s\}.
\end{equation}
This equality is identical to the column symmetric condition given in Definition \ref{defcolumnsym}.
\hfill\qed
\end{lemma}

The next lemma indicates that the ACWD is invariant 
after column shuffling.
\begin{lemma}[ACWD of a column shuffled ensemble]
\label{invariant}
For any $\ve s \in F_2^m$, $w \in [0,n]$, the equality
  \begin{equation}
    \tilde A_w^{(\Gamma_c(\Gen))}(\ve s) 
    = \tilde A_w^{(\Gen)}(\ve s)
  \end{equation}
holds. \\
(Proof)
From the equality
\begin{equation}
\#\{H\in \Gamma_c(\Gen): H \ve z^t = \ve s \}
=  \#\{H'\in \Gen: H' \ve z^t = \ve s \} n!,
\end{equation}
we have 
\begin{eqnarray} \nonumber
  \tilde A_w^{(\Gamma_c(\Gen))}(\ve s) 
  &=& \sum_{\ve z \in Z^{(n,w)}} \frac{\#\{H\in \Gamma_c(\Gen): 
    H \ve z^t = \ve s \}}{\# \Gamma_c(\Gen)} \\ \nonumber
  &=& \sum_{\ve z \in Z^{(n,w)}} \frac{
    \#\{H'\in \Gen: H' \ve z^t = \ve s \} n!}{n! \# \Gen}  \\
  &=& \tilde A_w^{(\Gen)}(\ve s).
\end{eqnarray}
\hfill\qed
\end{lemma}

\begin{example}
Let $H^*$ be an $m^* \times n^*$ binary matrix and
$M(H^*,\nu)$ be an $m \times n$ binary matrix with the form:
\begin{equation}
  M(H^*,\nu) \defeq
  \left(
  \begin{array}{cccc}
    H^* &     & &  \\
        & H^* & &  \\
        &     & \ddots &  \\
        &     &  & H^* \\
  \end{array}
  \right),
\end{equation}
where $\nu$ is a positive integer and $m = \nu m^*, n = \nu n^*$. 
The $\nu$-copies of the matrix $H^*$ are placed on 
the main diagonal of $M(H^*,\nu)$ and the remaining elements in $M(H^*,\nu)$ is 
set to be zero. The coset weight distribution corresponding to $M(H^*,\nu)$
is given by
\begin{equation} \label{diagonal}
  A_w(M(H^*,\nu),\ve s) = \left[\prod_{i=1}^\nu A_w(H^*,\ve s_i) \right]_w,
\end{equation}
where $\ve s = (\ve s_1,\ve s_2,\ldots, \ve s_\nu)$ and $\ve s_i \in F_2^{m/\nu} (i \in [1,\nu])$.
This equation means that,
if the coset weight distribution corresponding to $H^*$ is known, the coset weight distribution 
$A_w(M(H^*,\nu),\ve s)$ can be evaluated by using the generating function method.
From (\ref{diagonal}) and Lemma \ref{invariant}, we have the ACWD 
of the column shuffled ensemble $\Gamma_c(\Een(M(H^*,\nu)))$:
\begin{equation} 
  \tilde A_w(\ve s) = \left[\prod_{i=1}^\nu A_w(H^*,\ve s_i) \right]_w.
\end{equation}
\hfill\qed
\end{example}

\subsection{Row shuffled ensemble}
Assume that $\Gen$ (column size $n$ and row size $m$)
 is an ensemble which is not row symmetric.
The row shuffled ensemble $\Gamma_r(\Gen)$ is defined by
\begin{equation}
\Gamma_r(\Gen) \defeq \{\eta(H): H \in \Gen,\eta \in \psi_m \},
\end{equation} 
where $\psi_m$ is the set of all row permutations for $m$ rows.
Thus, $\#\Gamma_r(\Gen) = m! \#\Gen$.
The row shuffling converts an ensemble into a row symmetric one.
\begin{lemma}[Symmetric property of a row shuffled ensemble]
  \label{rowsynlemma}
  $\Gamma_r(\Gen)$ is row symmetric.
\hfill\qed
\end{lemma}
Since the proof of the lemma is almost the same as the proof of Lemma \ref{invariant},
it is omitted. 
The next lemma gives the relation between
the ACWD of an ensemble and that of the row shuffled one.
\begin{lemma}
  [ACWD of a row shuffled ensemble]
  \label{acwdrow}
  The ACWD of a row shuffled ensemble is expressed by
  \begin{equation}
    \tilde B_w^{(\Gamma_r(\Gen))}(\sigma) 
    = \sum_{\ve s \in Z^{(m,\sigma)}} \frac{1}{{m \choose \sigma}}
    \tilde A_w^{(\Gen)}(\ve s)
  \end{equation}
for $\sigma \in [0,m]$ and $w \in [0,n]$.  \\
(Proof)
The ACWD of a row shuffled ensemble can be derived in the 
following way:
\begin{eqnarray} \nonumber
  \tilde A_w^{(\Gamma_r(\Gen))}(\ve s) 
  \hspace{-3mm}
  &=& 
  \hspace{-3mm}
  \sum_{\ve z \in Z^{(n,w)}} \frac{
    \#\{H\in \Gamma_r(\Gen): H \ve z^t = \ve s \}}
  { \# \Gamma_r(\Gen)} \\ \nonumber
  &=& 
  \hspace{-3mm}
  \sum_{\ve z \in Z^{(n,w)}} \frac{
    \#\bigcup_{\eta \in \psi_m}\{H\in \Gen: H \ve z^t = \eta(\ve s) \}}
  { m! \# \Gen}  \\ \nonumber
  &=& 
  \hspace{-3mm}
  \sum_{\ve z \in Z^{(n,w)}} \sum_{\ve s' \in Z^{(m,|\ve s|)}} \frac{
    \#\{H\in \Gen: H \ve z^t = \ve s' \}} 
  { m! \# \Gen} \\ \nonumber
  &\times& \hspace{-3mm}(m-|\ve s|)! |\ve s|! \\ \nonumber
  &=& 
  \hspace{-3mm}
  \sum_{\ve s' \in Z^{(m,|\ve s|)}} 
  \frac{1}{{m \choose |\ve s|}}
  \sum_{\ve z \in Z^{(n,w)}}   
  \frac{
    \#\{H\in \Gen: H \ve z^t = \ve s' \}}
  {  \# \Gen} \\
  &=& 
  \hspace{-3mm}
  \sum_{\ve s' \in Z^{(m,|\ve s|)}} \frac{1}{{m \choose |\ve s|}}
    \tilde A_w^{(\Gen)}(\ve s').
\end{eqnarray}
In the above derivation, the second equality follows from
the relation $\eta(H)\ve z^t = \eta(H\ve z^t)$.
By replacing $|\ve s|$ by $\sigma$, we have the claim of the lemma.
\hfill\qed
\end{lemma}

\begin{example}
Let $\Gen$ be the Gallager ensemble with column size $n$ and row size $m$. 
We here consider 
the row shuffled ensemble $\Gamma_r(\Gen)$ and its ACWD. From Lemma \ref{acwdrow} and
the ACWD of the Gallager ensemble (\ref{gallagerACWD}), it is easy to derive the
ACWD of $\Gamma_c(\Gen)$:
\begin{eqnarray} \nonumber
  \tilde B_w(\sigma) &=& \frac{1}{{n \choose w}^{j-1} {m \choose \sigma} }
  \sum_{b_1 = 0}^{m/j} \cdots \sum_{b_j = 0}^{m/j}
  {\sigma \choose b_1,b_2,\ldots,b_j} \\
  &\times& \hspace{-3mm} \prod_{i=1}^j 
  \left[
    \alpha_k^{(m/j)-b_i}(x) \beta_k^{b_i}(x)
  \right]_{w},
\end{eqnarray}
where ${\sigma \choose b_1,b_2,\ldots,b_j}$ denotes a multinomial.
\hfill\qed
\end{example}

\section{Combined Ensembles}
\label{sec_combined}
In this section, we discuss ensembles which are composed of several simpler ensembles.
We call such an ensemble a combined ensemble. Two classes of 
combined ensembles,
which are called stacked ensembles and concatenated ensembles,
are bases of a complex combined ensemble.

\subsection{Stacked ensemble}
Let $\Aen$ and $\Ben$ be ensembles 
of $m_a \times n$ and $m_b \times n$ binary matrices, respectively.
The stacked ensemble based on $\Aen$ and $\Ben$ is defined as follows.
\begin{definition}[Stacked ensemble]
The stacked ensemble $\Aen/\Ben$ is defined by
\begin{equation}
  \Aen/\Ben \defeq \left\{
  \left[
  \begin{array}{c}
    H_a \\
    H_b
  \end{array}
  \right]
  : H_a \in \Aen, H_b \in \Ben  \right\},
\end{equation}
where each matrix in $\Aen/\Ben$ associates with probability
$1/(\#\Aen\#\Ben)$.
\hfill\qed
\end{definition}
The ensembles $\Aen$ and $\Ben$ are called {\em component ensembles}
of the stacked ensemble $\Aen/\Ben$.
In the following, 
for simplicity, we shall denote a stacked matrix as $[H_a/H_b]$.
The definition of the stack ensemble can be naturally extended to the case
where the number of component ensembles is more than two. 
For example, we define
$
  \Aen/\Ben/\Cen \defeq (\Aen/\Ben)/\Cen.
$

The next theorem shows the relation between the ACWD of a stacked ensemble
and the ACWDs of its component ensembles.
\begin{theorem}[ACWD of a stacked ensemble]
\label{th-stacked}
If the ensemble $\Ben$ are column symmetric, 
the ACWD of the stacked ensemble $\Aen / \Ben$
is given by
\begin{equation}
  \tilde A^{(\Aen / \Ben)}_{w}(\ve s)
  =
  \frac{1}{{n \choose w}}
  \tilde A^{(\Aen)}_{w}(\ve s_a)  \tilde A^{(\Ben)}_{w}(\ve s_b),
\end{equation}
where $\ve s = (\ve s_a, \ve s_b), \ve s_a \in F_2^{m_a}, 
\ve s_b \in F_2^{m_b}, w \in [0,n]$.\\
(Proof)
From the definition of the stacked ensemble, we have
\begin{eqnarray} \nonumber
  \tilde A^{(\Aen / \Ben)}_{w}(\ve s) \hspace{-3mm}&=& 
  \hspace{-6mm}
  \sum_{\ve z \in Z^{(n,w)}}
  \hspace{-3mm}
  \frac{\#\{H\in \Aen / \Ben:
    H \ve z^t = \ve s \}}{\#(\Aen / \Ben)}\\ \nonumber
  &=& 
  \hspace{-6mm}
  \sum_{\ve z \in Z^{(n,w)}} 
  \hspace{-3mm}
  \frac{\#\{[H_a/H_b] \in \Aen / \Ben:
    [H_a/H_b] \ve z^t = [\ve s_a/\ve s_b]  \}}{\#(\Aen / \Ben)},
\end{eqnarray}
where $\ve s_a$ and $\ve s_b$ are binary $m_a$ and $m_b$ tuples
satisfying $\ve s = (\ve s_a, \ve s_b)$.
We can further rewrite the above equation into the following form:
\begin{eqnarray} \nonumber
  \tilde A^{(\Aen / \Ben)}_{w}(\ve s) 
  &=& 
  \sum_{\ve z \in Z^{(n,w)}} \frac{
    \#\{H_a\in \Aen: H_a \ve z^t  = \ve s_a  \}
  }{\# \Aen }\\ \nonumber
  &\times& \hspace{-3mm} \frac{\#\{H_b\in \Ben: H_b \ve z^t  = \ve s_b  \}}{\# \Ben} \\  \nonumber
  &=& 
  \tilde A^{(\Aen)}_{w}(\ve s_a)
  \frac{
    \#\{H_b\in \Ben: H_b \ve z^t  = \ve s_b  \}
    } {
    \# \Ben
    }    {n \choose w} \frac{1}{{n \choose w}} \\ 
  &=& 
   \frac{1}{{n \choose w}}
   \tilde A^{(\Aen)}_{w}(\ve s_a)  \tilde A^{(\Ben)}_{w}(\ve s_b).
\end{eqnarray}
To derive the third equality from the second equality, 
the column symmetric property of $\Ben$ is used.
\hfill\qed
\end{theorem}

From the proof of Theorem \ref{th-stacked}, it is evident that
$\Aen/\Ben$ is column symmetric   
if both $\Aen$ and $\Ben$ are column symmetric. However, in general,
$\Aen/\Ben$ is not necessarily row symmetric 
even if both $\Aen$ and $\Ben$ are row symmetric.
The row symmetric property can be achieved by using row shuffling.
\begin{corollary}[Stacked ensemble with row shuffling]
  \label{stackacwd}
  If two ensembles $\Aen$ and $\Ben$ are column and row symmetric,
  then $\Gamma_r(\Aen / \Ben)$ is also  column and row symmetric.
  The ACWD of $\Gamma_r(\Aen / \Ben)$ is given by
  \begin{eqnarray} \nonumber
    \tilde B^{(\Gamma_r(\Aen / \Ben))}_{w}(\sigma) \hspace{-3mm} &=& \hspace{-3mm}
    \frac {1}{{n \choose w}{m \choose \sigma}}
    \sum_{\sigma_a = \max\{0,\sigma - m_b\}}^{\min\{\sigma,m_a\}}
    {m_a \choose \sigma_a}
    {m_b \choose \sigma-\sigma_a} \\
    &\times& \hspace{-3mm} \tilde B^{(\Aen)}_{w}(\sigma_a)
    \tilde B^{(\Ben)}_{w}(\sigma - \sigma_a)
  \end{eqnarray}
  for $\sigma \in [0,m], w \in [0,n]$, where $m \defeq m_a + m_b$. \\
(Proof) It is evident that
$\Gamma_r(\Aen / \Ben)$ is column and row symmetric. We focus on the 
proof of the ACWD. From Lemma \ref{acwdrow} and Theorem \ref{th-stacked},  
we obtain the following:
\[
    \tilde B_w^{(\Gamma_r(\Aen / \Ben))}(\sigma) 
    = \sum_{\ve s \in Z^{(m,\sigma)}} \frac{1}
    {{m \choose \sigma}}
    \tilde A_w^{(\Aen / \Ben)}(\ve s) \hspace{20mm}
\]
\vspace{-3mm}
\begin{eqnarray} \nonumber
    &=& \Theta \sum_{\ve s \in Z^{(m,\sigma)}} 
  \tilde A^{(\Aen)}_{w}(\ve s_a)  \tilde A^{(\Ben)}_{w}(\ve s_b) \\ \nonumber
    &=&  \Theta \hspace{-8mm}
    \sum_{\sigma_a = \max\{0,\sigma - m_b\}}^{\min\{\sigma,m_a\}}
    \sum_{\ve s_a' \in Z^{(m_a,\sigma_a)}}
    \sum_{\ve s_b' \in Z^{(m_b,\sigma-\sigma_a)}}
    \hspace{-4mm}
    \tilde A^{(\Aen)}_{w}(\ve s_a')  \tilde A^{(\Ben)}_{w}(\ve s_b'),
\end{eqnarray}
where 
\begin{equation}
  \Theta = 
    \frac{1}{{m \choose \sigma}{n \choose w}}.
\end{equation}
Applying the row symmetric property to the above equality,
we have the claim of the lemma.
\hfill\qed
\end{corollary}
This corollary can be easily extended to more general cases.
Let $\Aen_i (i \in [1,s])$ be a column and a row symmetric ensemble with row size $m_i$
and column size $n$.
The ACWD of $\Gamma_r(\Aen_1 / \Aen_2/ \cdots / \Aen_s)$ is given by
\begin{eqnarray} \nonumber
  \hspace{-5mm}
  \tilde B_w^{(\Gamma_r(\Aen_1 / \Aen_2/ \cdots / \Aen_s))}(\sigma) 
  \hspace{-3mm}
  &=& 
  \hspace{-3mm}  
  \frac{1}{ {m \choose \sigma}{n \choose w}^{(s-1)}} \\   \label{extendedcol}
  &\times& \hspace{-8mm} 
  \sum_{\stackrel{(\sigma_1,\ldots, \sigma_s),\sigma_i \ge 0}{\sigma = \sigma_1 +\cdots + \sigma_s}} 
\prod_{i=1}^s {m_i \choose \sigma_i}
\tilde B_w^{(\Aen_i)}(\sigma_i)
\end{eqnarray}
for $w \in [0,n]$ and $\sigma \in [0,m]$ where $m \defeq m_1+\ldots +m_s$.
Since the proof of (\ref{extendedcol}) is 
almost the same as that of Corollary \ref{stackacwd},
it is omitted.

\subsection{Concatenated ensemble}
Let $\Aen$ and $\Ben$ be  ensembles of $m \times n_a$ 
and $m \times n_b$ binary matrices, respectively.
The concatenated ensemble is defined as follows.
\begin{definition}[Concatenated ensemble]
The concatenated ensemble $\Aen\circ \Ben$ is defined by
\begin{equation}
  \Aen\circ \Ben \defeq \{[H_a H_b]: H_a \in \Aen, H_b \in \Ben  \},
\end{equation}
where a matrix $H$ in $\Aen\circ \Ben$ has the probability assignment
$P(H) = 1/(\#\Aen \#\Ben)$.
\end{definition}
A concatenated ensemble based on more than two ensembles can be defined 
as well as the case of the stacked ensemble.
For example, 
we define $\Aen \circ \Ben \circ \Cen \defeq (\Aen \circ \Ben) \circ \Cen$.

The next theorem presents the ACWD of a concatenated ensemble.
\begin{theorem}[ACWD of a concatenated ensemble]
  \label{th-concat}
  The ACWD of a concatenated ensemble $\tilde A^{(\Aen \circ \Ben)}_{w}(\ve s)$
  is given by
\begin{equation}
  \label{concatACWD}
  \tilde A^{(\Aen \circ \Ben)}_{w}(\ve s)
  =
  \sum_{w_a=\max\{0,w-n_b \}}^{\min\{w,n_a\}}
  \sum_{\ve s_a \in F_2^m}
  \tilde A^{(\Aen)}_{w_a}(\ve s_a)  \tilde A^{(\Ben)}_{w-w_a}(\ve s + \ve s_a)
\end{equation}
for $\ve s \in F_2^m$ and $w \in [0,n]$, where $n \defeq n_a + n_b$. \\
(Proof)
Let 
\begin{eqnarray}
  L_a(\ve s_1,\ve z_1) &=& \#\{H_a: H_a \in \Aen,H_a  \ve z_1^t = \ve s_1 \} \\
  L_b(\ve s_2,\ve z_2) &=& \#\{H_b: H_b \in \Ben,H_b \ve z_2^t = \ve s_2  \}
\end{eqnarray}
for $\ve z_1 \in F_2^{n_a}$, $\ve z_2 \in F_2^{n_b}$, $\ve s_1, \ve s_2 \in F_2^m$.
For any $H_a \in \Aen$, $H_b \in \Ben$ and $\ve z \in F_2^{n}$,  
the equality $\ve s = [H_a H_b] \ve z^t = H_a \ve z_a^t + H_b \ve z_b^t$
holds where $\ve z = (\ve z_a,\ve z_b)$. This equality leads to 
the following relation:
\begin{equation} \label{lalb}
  \#\{H\in \Aen \circ \Ben: H \ve z^t = \ve s \} = 
  \sum_{\ve s_a \in F_2^m}L_a(\ve s_a,\ve z_a) L_b(\ve s+\ve s_a,\ve z_b).
\end{equation}
By applying (\ref{lalb}) to the definition of the ACWD of the concatenated ensemble,
we have the claim of the theorem as follows.
\begin{eqnarray} \nonumber
  \tilde A^{(\Aen \circ \Ben)}_{w}(\ve s) 
  \hspace{-3mm}
  &=& 
  \hspace{-5mm}
  \sum_{\ve z \in Z^{(n,w)}} \frac{\#\{H\in \Aen \circ \Ben: H \ve z^t = \ve s \}}
  {\# \Aen \circ \Ben}\\ \nonumber
  &=& 
  \hspace{-5mm}
  \sum_{\ve z \in Z^{(n,w)}} 
  \frac{\sum_{\ve s_a \in F_2^m}L_a(\ve s_a,\ve z_a) L_b(\ve s+\ve s_a,\ve z_b)}
  {\# \Aen  \times \# \Ben}\\  \nonumber
  &=& 
  \hspace{-5mm}
\sum_{w_a=\max\{0,w-n_b \}}^{\min\{w,n_a\}}
\sum_{\ve s_a \in F_2^m}
\sum_{\ve z_a \in Z^{(n_a,w_a)}}
\frac{L_a(\ve s_a,\ve z_a)}{\#\Aen}  \\  \nonumber
&\times&
\hspace{-5mm}
\sum_{\ve z_b \in Z^{(n_b,w-w_a)}}
\frac{L_b(\ve s + \ve s_a,\ve z_b)}{\#\Ben} \\  \nonumber
 &=&
 \hspace{-5mm}
  \sum_{w_a=\max\{0,w-n_b \}}^{\min\{w,n_a\}}
  \sum_{\ve s_a \in F_2^m}
  \tilde A^{(\Aen)}_{w_a}(\ve s_a)  \tilde A^{(\Ben)}_{w-w_a}(\ve s + \ve s_a).
\end{eqnarray}
\hfill\qed
\end{theorem}

\begin{remark}
  From the above theorem, we can see that
  the average weight distribution of a concatenated ensemble is given by
  \begin{equation}
    \tilde A^{(\Aen \circ \Ben)}_{w}(\ve 0)
    =
    \sum_{w_a=\max\{0,w-n_b \}}^{\min\{w,n_a\}}
    \sum_{\ve s_a \in F_2^m}
    \tilde A^{(\Aen)}_{w_a}(\ve s_a)  \tilde A^{(\Ben)}_{w-w_a}(\ve s_a).
  \end{equation}
  This means that the ACWD of component ensembles plays a crucial role for deriving 
  the average weight distribution of a concatenated ensemble.
  \hfill\qed
\end{remark}

\begin{remark}
  In some cases, the weight corresponding to each component ensemble
  should be treated separately.
  For example, when some bits corresponding to a component ensemble 
  are punctured, analysis based on a split weight distribution 
  is required. 
  Such a distribution
  is discussed in \cite{shimo04} \cite{Ikegaya}.

  In such a case, it is useful to define a split ACWD.
  We here do not go into the details such as 
  the definition of a split weight distribution but
  just see the split ACWD version of Theorem \ref{th-concat}:
  \begin{equation}
    \tilde A^{(\Aen \circ \Ben)}_{w_1,w_2}(\ve s)
    = \sum_{\ve s_a \in F_2^m}
    \tilde A^{(\Aen)}_{w_1}(\ve s_a)  \tilde A^{(\Ben)}_{w_2}(\ve s + \ve s_a).
  \end{equation}
  \hfill\qed
\end{remark}

The exponential number of summands  (in terms of $m$) in (\ref{concatACWD})
can be reduced when $\Aen$ and $\Ben$ are row symmetric.
The following corollary deals with such a case.
\begin{corollary}[Case of row symmetric component ensembles]
\label{concatcoro}
If both $\Aen$ and $\Ben$ are row symmetric, then
$\Aen \circ \Ben$ is also row symmetric.
The ACWD of $\Aen \circ \Ben$ is given by
\begin{eqnarray} \nonumber
  \tilde B^{(\Aen \circ \Ben)}_{w}(\sigma)
  &=& 
  \sum_{w_a=\max\{0,w-n_b \}}^{\min\{w,n_a\}}
  \sum_{\mu_1 = 0}^{\sigma} \sum_{\mu_2 = 0}^{m-\sigma}
  {\sigma \choose \mu_1} {m-\sigma \choose \mu_2} \\   \label{rowsynacwd}
  &\times&
  \tilde B^{(\Aen)}_{w_a}(\mu_1+\mu_2)  
  \tilde B^{(\Ben)}_{w-w_a}(\sigma - \mu_1+\mu_2).
\end{eqnarray}
(Proof)
Suppose that $\ve s \in F_2^m$ is given.
Assume that two binary vectors $\ve s_a= (s_{a1},s_{a2},\ldots,s_{am})$ 
and $\ve s_b = (s_{b1},s_{b2},\ldots,s_{bm})$
satisfy $\ve s = \ve s_a + \ve s_b$.
We define two index sets in the following way:
\begin{eqnarray}
I_{1} &\defeq& \{i \in [1,m]:s_{ai} \ne s_{bi}\} \\
I_{2} &\defeq& \{i \in [1,m]:s_{ai} = s_{bi}\}.
\end{eqnarray}
Since $\ve s = \ve s_a + \ve s_b$ holds, the size of $I_1$ and $I_2$  must be
$|\ve s|$ and $m- |\ve s|$, respectively.

We here define $\mu_1$ and $\mu_2$ by
$\mu_1 \defeq \sum_{i \in I_1} |s_{ai}|, \mu_2 \defeq \sum_{i \in I_2} |s_{ai}|$.
Namely, $\mu_1$ is the Hamming weight of a vector 
$(s_{ai})_{i \in I_1}$ and $\mu_2$ is the Hamming weight of $(s_{ai})_{i \in I_2}$.
Note that the equalities $|\ve s_a| = \mu_1 + \mu_2$ and 
$|\ve s_b| = |\ve s| - \mu_1 + \mu_2$ holds 
where $\mu_1 \in [0,|\ve s|]$, $\mu_2 \in [0,m-|\ve s|]$.
Consider the size of the set which includes all $m$-tuples
corresponding to $\ve s$, $\mu_1$ and $\mu_2$:
\begin{eqnarray} \nonumber
  \phi(\ve s, \mu_1, \mu_2) \defeq \#\{\ve s_a \in F_2^m:
  |\ve s_a|= \mu_1+\mu_2, \\
  |\ve s + \ve s_a| = |\ve s| -\mu_1+\mu_2\}
\end{eqnarray}
where $\ve s \in F_2^m, \mu_1 \in [0,|\ve s|]$, $\mu_2 \in [0,m-|\ve s|]$.
Based on a simple combinatorial argument,
we have
\begin{equation}
  \phi(\ve s, \mu_1, \mu_2) = {|\ve s| \choose \mu_1} {m-|\ve s| \choose \mu_2}.
\end{equation}

From Theorem \ref{th-concat} and the above discussion, 
the ACWD of $\Aen \circ \Ben$ can be written into the 
following form:
\begin{eqnarray} \nonumber
  \tilde A^{(\Aen \circ \Ben)}_{w}(\ve s)
  \hspace{-2mm}
  &=&
  \hspace{-8mm}
  \sum_{w_a=\max\{0,w-n_b \}}^{\min\{w,n_a\}}
  \sum_{\ve s_a \in F_2^m}
  \tilde B^{(\Aen)}_{w_a}(|\ve s_a|)  
  \tilde B^{(\Ben)}_{w-w_a}(|\ve s + \ve s_a|) \\ \nonumber
  &=&
  \hspace{-8mm}
  \sum_{w_a=\max\{0,w-n_b \}}^{\min\{w,n_a\}}
  \sum_{\mu_1 = 0}^{|\ve s|}
  \sum_{\mu_2 = 0}^{m-|\ve s|}
  \phi(\ve s, \mu_1, \mu_2) \\ \nonumber
  &\times&
  \hspace{-3mm}
  \tilde B^{(\Aen)}_{w_a}(\mu_1+\mu_2)  
  \tilde B^{(\Ben)}_{w-w_a}(|\ve s|-\mu_1+\mu_2)  \\ \nonumber
  &=&
  \hspace{-8mm}
  \sum_{w_a=\max\{0,w-n_b \}}^{\min\{w,n_a\}}
  \sum_{\mu_1 = 0}^{|\ve s|}
  \sum_{\mu_2 = 0}^{m-|\ve s|}
  {|\ve s| \choose \mu_1} {m-|\ve s| \choose \mu_2} \\
  &\times&
  \hspace{-3mm}
  \tilde B^{(\Aen)}_{w_a}(\mu_1+\mu_2)  
  \tilde B^{(\Ben)}_{w-w_a}(|\ve s|-\mu_1+\mu_2) 
\end{eqnarray}
for any $\ve s \in F_2^m, w \in [0,n]$.
We can see that the right hand side of the above equation 
depends only on the weight
of $\ve s$. This means that $\Aen \circ \Ben$ is row symmetric.
Replacing $|\ve s|$ by $\sigma$, we have the claim of the corollary.
\hfill\qed
\end{corollary}

\begin{remark}
  If $\Aen$ and $\Ben$ are both row symmetric,
  then the average weight distribution of the concatenated 
  ensemble $\Aen \circ \Ben$ can be derived from (\ref{rowsynacwd}),
  which is given by
 \begin{equation}
   \tilde B^{(\Aen \circ \Ben)}_{w}(0)
   =
   \sum_{w_a=\max\{0,w-n_b \}}^{\min\{w,n_a\}}
   \sum_{\nu = 0}^m {m \choose \nu}
   \tilde B^{(\Aen)}_{w_a}(\nu)  \tilde B^{(\Ben)}_{w - w_a}(\nu).
 \end{equation}
 Related results have been presented in \cite{dru} for IRA code ensembles
 and \cite{shimo04} \cite{Ikegaya} for tripartite graph ensembles.
\hfill\qed
\end{remark}

\begin{example}
  Let $\Cen_a$ be the 
  $(j = 2, k=4)$-regular bipartite ensemble with column size $n = 6$ and row size $m = 3$
  and $\Cen_b$ be the 
  $(j = 1, k=2)$-regular bipartite ensemble with the same column and row size. 
  Table \ref{compentacwd}
  presents the ACWD $\tilde B_w(\sigma)$ of $\Cen_a$ and $\Cen_b$ which are 
  evaluated using (\ref{acwdbipartite}). The ACWDs of the stacked ensemble 
  $\Cen_a/\Cen_b$ and the concatenated ensemble $\Cen_a \circ \Cen_b$ are
  derived by Corollary \ref{stackacwd} and Corollary \ref{concatacwd}, respectively.
  They are shown in Tables \ref{tblstacked} and \ref{tblconcat}.
  \hfill\qed
\end{example}

\begin{table}
\caption{The ACWD of regular bipartite ensembles}
\label{compentacwd}
\centering{(a) $\Cen_a$ $(j = 2, k=4)$}
\[
\begin{array}{c|ccccccc}
\hline
\hline
\sigma \backslash w & 0 & 1 & 2 & 3 & 4 & 5 & 6 \\
\hline
0&1&\frac{18}{11}&\frac{37}{11}&\frac{60}{11}&\frac{37}{11}&\frac{18}{11}&1 \\
1&0&0&0&0&0&0&0 \\
2&0&\frac{16}{11}&\frac{128}{33}&\frac{160}{33}&\frac{128}{33}&\frac{16}{11}&0 \\
3&0&0&0&0&0&0&0 \\
\hline
\end{array}
\]
\centering{(b) $\Cen_b$ $(j = 1, k=2)$}
\[
\begin{array}{c|ccccccc}
\hline
\hline
\sigma \backslash w & 0 & 1 & 2 & 3 & 4 & 5 & 6 \\
\hline
0&1&0&3&0&3&0&1 \\
1&0&2&0&4&0&2&0 \\
2&0&0&4&0&4&0&0 \\
3&0&0&0&8&0&0&0 \\
\hline
\end{array}
\]
\centering{(column size 6, row size 3)}
\end{table}

\begin{table}
\caption{The ACWD of the stacked ensemble $\Cen_a/\Cen_b$}
\label{tblstacked}
\[
\begin{array}{c|ccccccc}
\hline
\hline
\sigma \backslash w & 0 & 1 & 2 & 3 & 4 & 5 & 6 \\
\hline
0&1&0&\frac{37}{55}&0&\frac{37}{55}&0&1 \\
1&0&\frac{3}{11}&0&\frac{6}{11}&0&\frac{3}{11}&0 \\
2&0&0&\frac{92}{275}&0&\frac{92}{275}&0&0 \\
3&0&\frac{12}{55}&0&\frac{6}{11}&0&\frac{12} {55}&0 \\
4&0&0&\frac{512}{825}&0&\frac{512}{825}&0&0 \\
5&0&0&0&\frac{32}{33}&0&0&0 \\
6&0&0&0&0&0&0&0 \\
\hline
\end{array}
\]
\centering{(column size 6, row size 6)}
\end{table}

\begin{table*}
\caption{The ACWD  of the concatenated ensemble $\Cen_a \circ \Cen_b$}
\label{tblconcat}
\[
  \begin{array}{c|ccccccccccccc}
    \hline
    \hline
    \sigma\backslash w & 0 & 1 & 2 & 3 & 4 & 5 & 6 & 7 & 8 & 9 & 10 & 11 & 12 \\
    \hline
    0&1&\frac{18}{11}&\frac{70}{11}&\frac{306}{11}&63&\frac{1084}{11}&\frac{1268}{11}&\frac{1084}{11}&63&\frac{306}{11}&\frac{70}{11}&\frac{18}{11}&1 \\[1mm]
    1&0&2&\frac{100}{11}&\frac{866}{33}&\frac{1984}{33}&\frac{3292}{33}&\frac{3880}{33}&\frac{3292}{33}&\frac{1984}{33}&\frac{866}{33}&\frac{100}{11}&2&0 \\[1mm]
    2&0&\frac{16}{11}&\frac{260}{33}&\frac{904}{33}&64&\frac{3272}{33}&\frac{3704}{33}&\frac{3272}{33}&64&\frac{904}{33}&\frac{260}{33}&\frac{16}{11}&0 \\[1mm]
    3&0&0&\frac{96}{11}&\frac{344}{11}&\frac{656}{11}&\frac{1064}{11}&\frac{1312}{11}&\frac{1064}{11}&\frac{656}{11}&\frac{344}{11}&\frac{96}{11}&0&0 \\[1mm]
    \hline
  \end{array}
\]
\centering{(column size 12, row size 3)}
\end{table*}

In general, a concatenated ensemble $\Aen \circ \Ben$ is not
column symmetric. 
The following corollary is a direct consequence of Corollary \ref{concatcoro} and
Lemma \ref{rowsynlemma}.
\begin{corollary}
\label{concatacwd}
If both $\Aen$ and $\Ben$ are row symmetric, then
the column shuffled ensemble $\Gamma_c(\Aen \circ \Ben)$ is
column and row symmetric.
The ACWD $\tilde B^{(\Gamma_c(\Aen \circ \Ben))}_{w}(\sigma)$ is 
the same as that given in  (\ref{rowsynacwd}).
\end{corollary}

\subsection{Combined ensembles}

In the previous subsections, we have seen that an
ensemble can be defined based on simpler ensembles.
A recursive use of stacking operation and concatenation 
operation yields a more complex ensemble.
The following definition characterize a combined ensemble constructed from
simpler ensembles.
\begin{definition}[Combined ensemble]
  Let $\Gen_1, \Gen_2, \ldots, \Gen_t$ be binary matrix ensembles.
  If an ensemble $\Een$ is obtained by combining $\Gen_1, \ldots, \Gen_t$ with
  stack operation, concatenation operation, column shuffling, and row shuffling,
  then $\Een$ is said to be a combined ensemble.
  Every ensemble $\Gen_i (i \in [1,t])$ is called a component ensemble of $\Een$.
  \hfill\qed
\end{definition}

\subsubsection{Type I combined ensembles}

In principle, the ACWD of a combined ensemble is evaluated 
by using Theorems \ref{th-stacked} and \ref{th-concat} but 
we focus on subclasses of combined ensembles whose 
ACWD can be represented by a simple form.
\begin{definition}[Type I combined ensemble]
  Let $\Een$ be a combined ensemble expressed as
  \begin{equation}
    \Een = \Aen_1 \circ \Aen_2 \circ \Aen_3 \circ \cdots \circ \Aen_s.
  \end{equation}
  The ensemble $\Aen_i (i \in [1,s] )$ has column size $n_i$ and row size $m$
  where $n \defeq n_1 + n_2 + \cdots + n_s$.
  Each $\Aen_i (i \in [1,s] )$ satisfies the following Conditions  C1 or C2,
  then the combined ensemble $\Een$ is said to be a type I combined ensemble:
  \begin{description}
  \item[(C1)]   $\Aen_i = \Gamma_r(\Ben_{i1}/\Ben_{i2}/\cdots /\Ben_{iu_i})$ holds where
    each $\Ben_{i\ell} (\ell \in [1,u_i])$ 
    is a column and row symmetric ensemble with column size $n_i$ and row size
    $m_{i\ell}$. Note that $m = m_{i1} + m_{i2} + \cdots + m_{i{u_i}}$ must be satisfied.
  \item[(C2)]   $\Aen_i = \Ben_{i1}$ holds 
    where $\Ben_{i1}$ is a row symmetric ensemble 
    with column size $n_i$ and row size $m$.
  \end{description}
  \hfill\qed
\end{definition}

\begin{example}
  Figure \ref{extypeI} presents a configuration of a type I combined ensemble.
   \begin{equation}
     \Een \defeq \Gamma_r(\Ben_{11}/\Ben_{12}/\Ben_{13}) \circ
     \Gamma_r(\Ben_{21}/\Ben_{22}/\Ben_{23}/\Ben_{24}) \circ \Ben_{31}
   \end{equation}
   Note that row shuffling operations are omitted in Fig.\ref{extypeI}.
  \hfill\qed
\begin{figure}[htbp]
  \begin{center}
    \begin{picture}(120,80)
      \put(0,0){\framebox(100,50)}
      \put(0,0){\framebox(40,15)}
      \put(0,15){\framebox(40,15)}
      \put(0,30){\framebox(40,20)}
      \put(40,0){\framebox(40,12)}
      \put(40,12){\framebox(40,12)}
      \put(40,24){\framebox(40,12)}
      \put(40,36){\framebox(40,14)}
      \put(10,35){$\Ben_{11}$}
      \put(10,20){$\Ben_{12}$}
      \put(10,5){$\Ben_{13}$}
      \put(53,40){$\Ben_{21}$}
      \put(53,27){$\Ben_{22}$}
      \put(53,15){$\Ben_{23}$}
      \put(53,3){$\Ben_{24}$}
      \put(83,20){$\Ben_{31}$}
      \put(10,55){$n_1$}
      \put(53,55){$n_2$}
      \put(83,55){$n_3$}
      \put(5,65){$u_1 = 3$}
      \put(47,65){$u_2 = 4$}
      \put(83,65){$u_3 = 1$}
    \end{picture}
  \end{center}
  \caption{Example of a type I combined ensemble}
  \label{extypeI}
\end{figure}
\end{example}

The following corollary shows that the ACWD of a type I combined ensemble 
can be obtained based on the recursive use of Corollary \ref{concatacwd}.
\begin{corollary}[ACWD of Type I combined ensemble]
\label{typeI}
  Let 
  \begin{equation}
    \Den \defeq \Aen_1 \circ \Aen_2 \circ \cdots \circ \Aen_{i-1},
  \end{equation}
  where $n_a \defeq \sum_{j=1}^{i-1} n_j$ and $n_b \defeq n_i$.
  The ACWD of $\Den \circ \Aen_i (i \in [2,s])$ is given by
\begin{eqnarray} \nonumber
  \tilde B^{(\Den \circ \Aen_i)}_{w}(\sigma)
  &=& 
  \sum_{w_a=\max\{0,w-n_b \}}^{\min\{w,n_a\}}
  \sum_{\mu_1 = 0}^{\sigma} \sum_{\mu_2 = 0}^{m-\sigma}
  {\sigma \choose \mu_1} {m-\sigma \choose \mu_2} \\   
  &\times&
  \tilde B^{(\Den)}_{w_a}(\mu_1+\mu_2)  
  \tilde B^{(\Aen_i)}_{w-w_a}(\sigma - \mu_1+\mu_2),
\end{eqnarray}
where $\tilde B^{(\Aen_i)}_{w}(\sigma)$ can be obtained by using 
(\ref{extendedcol})
if $\Aen_i = \Gamma_r(\Ben_{i1}/\Ben_{i2}/\cdots /\Ben_{iu_i})$. \\
(Proof) 
The claim of the corollary is a direct consequence of Corollary \ref{concatacwd}.
\hfill\qed
\end{corollary}

\begin{remark}
  Let $\Een$ is a type I combined ensemble constructed 
  based on the set of component ensembles $\{\Ben_{i\ell}\}$.
  Corollary \ref{typeI} implies that
  the ACWD $B_w^{(\Een)}(\sigma)$ 
  is expressed by a linear combination of the terms  which have the following form
  \begin{equation}
    \prod_{i= 1}^{s}
    \prod_{\ell = 1}^{u_i} \tilde B_{w_i}^{(\Ben_{i\ell})}(\sigma_{i\ell}),
  \end{equation}
  where $w_i \in [0,n_i]$ and $\sigma_{i\ell} \in [0,m_{i\ell}]$.
  A term corresponds to 4 parameters such as $s, u_i, w_i$ and $\sigma_{i \ell}$.
  Since the number of possible way to choose a set of parameters can be bounded by $n^4$,
  the number of the terms in the linear combination is upper bounded by $n^4$.
  \hfill\qed
\end{remark}

The proof of Corollary \ref{typeI} shows that (\ref{extendedcol}) and 
Corollary \ref{concatacwd} are key tools to evaluate the ACWD of a type I 
combined ensemble. Furthermore, we can observe that the number of terms grows
with polynomial order $O(n^4)$. This polynomial growth property simplifies 
the analysis on the asymptotic growth rate of the ACWD (discussed in the next section) 
of a type I combined ensemble.

\subsubsection{Type II combined ensembles}
A type I combined ensemble is very easy to analyze
because every $\Aen_i$ is row symmetric.
We consider another type of combined ensemble, which includes
ensembles of multi-edge type LDPC codes.
\begin{definition}[Type II combined ensemble]
  Let $\Een$ be a combined ensemble expressed as
  \begin{equation}
    \Een = \Aen_1 \circ \Aen_2 \circ \Aen_3 \circ \cdots \circ \Aen_s,
  \end{equation}
  and each ensemble $\Aen_i (i \in [1,s] )$ with column size $n_i$
  ($n = n_1 + n_2 +\cdots + n_s$) and row size $m$
  satisfies the following Conditions  C1' or C2',
  then the combined ensemble $\Een$ is said to be a type II combined ensemble.
  \begin{description}
  \item[(C1')]   $\Aen_i = \Ben_{i1}/\Ben_{i2}/\cdots /\Ben_{iu}$ holds where
    each $\Ben_{i\ell} (\ell \in [1,u])$ is 
    a column and row symmetric ensemble with column size $n_i$ and row size $m_\ell$.
    Note that $m = m_1 + m_2 + \cdots + m_\ell$ must be satisfied.
  \item[(C2')]   $\Aen_i = \Ben_{i1}$ holds 
    where $\Ben_{i1}$  is a row symmetric ensemble with column size $n_i$ and row size 
    $m$.
  \end{description}
  \hfill\qed
\end{definition}

\begin{example}
Figure \ref{extypeII} presents a configuration of a type II combined ensemble
\begin{equation}
  \Een \defeq (\Ben_{11}/\Ben_{12}/\Ben_{13}) \circ (\Ben_{21}/\Ben_{22}/\Ben_{23}) \circ \Ben_{31}.
\end{equation}
  \hfill\qed
\begin{figure}[htbp]
  \begin{center}
    \begin{picture}(120,80)
      \put(0,0){\framebox(100,50)}
      \put(0,0){\framebox(40,15)}
      \put(0,15){\framebox(40,15)}
      \put(0,30){\framebox(40,20)}
      \put(40,0){\framebox(40,15)}
      \put(40,15){\framebox(40,15)}
      \put(40,30){\framebox(40,20)}
      \put(10,35){$\Ben_{11}$}
      \put(10,20){$\Ben_{12}$}
      \put(10,5){$\Ben_{13}$}
      \put(53,35){$\Ben_{21}$}
      \put(53,20){$\Ben_{22}$}
      \put(53,5){$\Ben_{23}$}
      \put(83,20){$\Ben_{31}$}
      \put(10,55){$n_1$}
      \put(53,55){$n_2$}
      \put(83,55){$n_3$}
      \put(50,65){$u = 3$}
      \put(-20,35){$m_1$}
      \put(-20,20){$m_2$}
      \put(-20,5){$m_3$}
    \end{picture}
  \end{center}
  \caption{Example of a type II combined ensemble}
  \label{extypeII}
\end{figure}
\end{example}

Note that, in general,  $\Aen_i (i \in [1,s])$ is not row symmetric
when $\Aen_i = \Ben_{i1}/\Ben_{i2}/\cdots /\Ben_{iu}$. This fact
makes analysis more intricate than that of a type I combined ensemble.

Consider the case where $\Aen_i = \Ben_{i1}/\Ben_{i2}/\cdots /\Ben_{iu}$.
Each $\Ben_{i\ell} (\ell \in [1,u])$ is 
a column and row symmetric ensemble with column size $n$ and row size $m_\ell$.
From Theorem \ref{th-stacked}, we immediately have the ACWD of $\Aen_i$:
\begin{equation}
  \tilde A_w^{(\Aen_i)}(\ve s) = \frac{1}{{n \choose w}^{u-1}}
  \tilde A_w^{(\Ben_{i1})}(\ve s_1) \tilde A_w^{(\Ben_{i2})}(\ve s_2) \cdots
  \tilde A_w^{(\Ben_{iu})}(\ve s_u)
\end{equation}
where $\ve s = (\ve s_1,\ldots, \ve s_u)$ and $\ve s_\ell \in F_2^{m_\ell} (\ell \in [1,u])$.
From the assumption that $\Ben_{i\ell}$ is row symmetric, it is evident that
the ACWD depends on $(|\ve s_1|, |\ve s_2|,\ldots, |\ve s_u|)$;
namely 
\begin{equation} \label{separate}
  \tilde A_w^{(\Aen_i)}(\ve s) = 
\frac{1}{{n \choose w}^{u-1}} \prod_{\ell=1}^u \tilde B_w^{(\Ben_{i\ell})}(|\ve s_{\ell}|).
\end{equation}
In order to evaluate the ACWD of a type II combined ensemble,
it is natural to define the following variant of the ACWD, which is called
the ACWD {\em in split syndrome form}.
\begin{definition}
  If $\Aen_i = \Ben_{i1}/\Ben_{i2}/\cdots /\Ben_{iu}$ holds, then
  $\tilde C_w^{(\Aen_i)}(\sigma_1,\ldots, \sigma_u)$ is defined by
\begin{equation}
  \tilde C_w^{(\Aen_i)}(\sigma_1,\ldots, \sigma_u) \defeq 
  \frac{1}{{n \choose w}^{u-1}} \prod_{\ell=1}^u \tilde B_w^{(\Ben_{i\ell})}(\sigma_\ell),
\end{equation}
where $\sigma_\ell \in [0,m_\ell]$ for each $\ell \in [1,u]$.
If $\Aen_i = \Ben_{i1}$ then,  let
\begin{equation}
  \tilde C_w^{(\Aen_i)}(\sigma_1,\ldots, \sigma_u) 
\defeq \tilde B_w^{(\Ben_{i1})}(\sigma_1+\cdots+\sigma_u).
\end{equation}
\hfill\qed
\end{definition}

From this definition and (\ref{separate}), we see that the following equation
\begin{equation}
  \tilde A_w^{(\Aen_i)}(\ve s) = 
  \tilde C_w^{(\Aen_i)}(|\ve s_1|,\ldots, |\ve s_u|)  
\end{equation}
holds for each $i \in [1,s]$. 
By using the ACWD in split syndrome form, we can handle the ACWD of
a type II combined ensemble. The following corollary explains the details.
\begin{corollary}[ACWD of Type II combined ensemble]
\label{col_type2}
  Let 
  \begin{equation}
    \Den \defeq \Aen_1 \circ \Aen_2 \circ \cdots \circ \Aen_{i-1},
  \end{equation}
  where $n_a \defeq \sum_{j=1}^{i-1} n_j$ and $n_b \defeq n_i$.
  The ACWD of $\Den \circ \Aen_i (i \in [2,s])$ is given by
\[
  \tilde C_w^{(\Den \circ \Aen_i)}(\sigma_1,\ldots, \sigma_u)
  = \hspace{-10mm} \sum_{w_a=\max\{0,w-n_b \}}^{\min\{w,n_a\}}
  \sum_{p_1 = 0}^{\sigma_1} \sum_{q_1 = 0}^{m_1-\sigma_1}
  \cdots \sum_{p_u = 0}^{\sigma_u} \sum_{q_u = 0}^{m_u-\sigma_u}
\]
\vspace{-4mm}
\begin{eqnarray} \nonumber
  &\times& \hspace{-3mm} \prod_{j=1}^u {\sigma_j \choose p_j} {m_j-\sigma_j \choose q_j} 
 \tilde C_{w_a}^{(\Den)}(p_1+q_1,\ldots, p_u+q_u) \\
  &\times& \hspace{-3mm} C_{w-w_a}^{(\Aen_{i})}(\sigma_1- p_1+q_1,\ldots, \sigma_u - p_u+q_u).
\end{eqnarray}
(Proof) 
From Theorem \ref{th-concat}, we have 
\begin{eqnarray} \nonumber
  \tilde A_w^{(\Den \circ \Aen_i)}(\ve s) 
  \hspace{-3mm}
  &=& 
  \hspace{-5mm}
  \sum_{\ve s_a \in F_2^m} 
  \sum_{w_a=\max\{0,w-n_b \}}^{\min\{w,n_a\}} \tilde A_{w_a}^{(\Den)}(\ve s_a) 
  \tilde A_{w - w_a}^{(\Aen_i)}(\ve s +\ve s_a) \\ \nonumber
  &=& 
  \hspace{-5mm}
  \sum_{\ve s_a \in F_2^m} 
  \sum_{w_a=\max\{0,w-n_b \}}^{\min\{w,n_a\}} 
  \tilde C_{w_a}^{(\Den)}(|\ve s_{a1}|,\ldots,|\ve s_{au}|) \\
  &\times&
  \hspace{-3mm}
  \tilde C_{w - w_a}^{(\Aen_i)}(|\ve s_1 +\ve s_{a1}|, \ldots,|\ve s_u +\ve s_{au}|),
\end{eqnarray}
where $\ve s = (\ve s_1, \ldots, \ve s_u)$ and $\ve s_a = (\ve s_{a1}, \ldots, \ve s_{au})$.
Based on the arguments similar to those used in the proof of Corollary \ref{concatcoro},
$\tilde A_w^{(\Den \circ \Aen_i)}(\ve s)$ can be transformed in such a way:
\[
  \tilde A_w^{(\Den \circ \Aen_i)}(\ve s) \hspace{70mm}
\]
\vspace{-4mm}
\begin{eqnarray} \nonumber
  &=& 
  \hspace{-8mm}
  \sum_{w_a=\max\{0,w-n_b \}}^{\min\{w,n_a\}} 
  \sum_{\ve s_{a1} \in F_2^{m_1}} 
  \cdots
  \sum_{\ve s_{au} \in F_2^{m_u}} 
  \tilde C_{w_a}^{(\Den)}(|\ve s_{a1}|,\ldots,|\ve s_{au}|) \\
  &\times& 
  \hspace{-3mm}
  \tilde C_{w - w_a}^{(\Aen_i)}(|\ve s_1 +\ve s_{a1}|, \ldots,|\ve s_u +\ve s_{au}|) \\ 
  \nonumber
  &=&
  \hspace{-8mm}
  \sum_{w_a=\max\{0,w-n_b \}}^{\min\{w,n_a\}}
  \sum_{p_1 = 0}^{|\ve s_1|} \sum_{q_1 = 0}^{m_1-|\ve s_1|}
  \cdots \sum_{p_u = 0}^{|\ve s_u|} \sum_{q_u = 0}^{m_u- |\ve s_u|}
  \prod_{j=1}^u \phi(\ve s_j, p_j, q_j) \\ \nonumber
  &\times& 
  \hspace{-3mm}
  \tilde C_{w_a}^{(\Den)}(p_1+q_1,\ldots, p_u+q_u) \\
  &\times& 
  \hspace{-3mm}
  C_{w-w_a}^{(\Aen_{i})}(|\ve s_1|- p_1+q_1,\ldots, |\ve s_u| - p_u+q_u).
\end{eqnarray}
By replacing $|\ve s_j|$ by $\sigma_j ( \sigma_j\in [0,m_j])$ 
for each $j (j \in [1,u])$,
we obtain the claim of the corollary.
\hfill\qed
\end{corollary}

\section{Asymptotic Growth Rate of the ACWD}
\label{sec_asymptotic}
The asymptotic growth rate (abbreviated as the AGR) of an average weight distribution of an ensemble
reflects the asymptotic behavior of a code included in the ensemble.
Thus, the AGR of an average weight distribution 
has been extensively studied in \cite{LS02} \cite{LS03} \cite{MV04} \cite{dru}.
In this section, we discuss the {\em asymptotic growth rate of the ACWD},
which means the asymptotic growth rate of $\tilde A_{w}(\ve s)$.
For simplicity, we only focus on the row symmetric ensembles in this section.

\subsection{AGR of a regular bipartite ensemble}
\begin{definition}[AGR of the ACWD]
Let $\{\Gen_n \}_{n>0}$ be a sequence of row symmetric 
ensembles. The ensemble  $\Gen_n$ has  column size $n$ and row size $(1-R)n$,
where $R (0 < R < 1)$ is called the {\em design rate} of the ensemble.
The asymptotic growth rate of the ACWD of $\{\Gen_n \}_{n>0}$ 
is defined by
\begin{equation}
  \tilde b_{\ell}(\eta) \defeq  \lim_{n\rightarrow \infty} \frac{1}{n} \log_2 
  \tilde B^{(\Gen_n)}_{n \ell}((1-R)n \eta),
\end{equation}
where  $0 \le \ell \le 1$ and $0 \le \eta \le 1$.
\hfill\qed
\end{definition}
Note that the above definition is consistent with the definition of 
the asymptotic growth rate 
of an average weight distribution discussed
in \cite{LS02} \cite{LS03} \cite{MV04} \cite{dru}.
Namely, $\tilde b_{\ell}(0)$ is exactly same as 
the asymptotic growth rate of an average weight distribution.

The saddle point method (Hayman method \cite{dru} \cite{modern}) is quite useful 
to estimate coefficients of a generating function and yields closed form 
formulas for asymptotic growth rate.
The next lemma presents the asymptotic growth rate of the sequence of 
the $(j,k)$-regular bipartite ensembles.
\begin{lemma}[AGR of bipartite ensembles]
  Let $\{\Gen_n \}_{n > 0}$ be 
  the sequence of the $(j,k)$-regular bipartite ensembles with design rate $R = 1-j/k$.
  The asymptotic growth rate of this sequence of ensembles is given by 
  \begin{eqnarray} \nonumber
    \tilde b_{\ell}(\eta)
    &=& \frac{j}{k} \left((1-\eta) \log_2 \alpha_k(r) + \eta \log_2 \beta_k(r) 
      -\ell k\log_2 r \right)  \\
    &-& \hspace{-3mm}(j-1) H(\ell),
  \end{eqnarray}
  where $r$ is the smallest positive root of
  \begin{equation}
     x \frac{(\alpha^{1-\eta}_k(x) \beta_k^\eta(x))'}{\alpha^{1-\eta}_k(x) \beta_k^\eta(x)}
     =\ell k
  \end{equation}
  and $H(x)$ is the binary entropy function defined by
  \begin{equation}
    H(x) \defeq -x \log_2 (x) - (1-x) \log_2 (1-x).
  \end{equation}
(Proof)
Let 
\begin{equation}
  f(x) \defeq \alpha_k(x)^{1-\eta} \beta_k(x)^{\eta}.
\end{equation}
From Lemma \ref{bipartite}, we have
\begin{equation}
  \tilde B_{\ell n}((1-R) n \eta) 
  = \frac{ 
    [f(x)^{(1-R)n}]_{\ell n j}
    }
  {{n j \choose \ell n j}}
  {n \choose \ell n}.
\end{equation}
The asymptotic growth rate $\tilde b_{\ell}(\eta)$ can be 
rewritten as follows:
\begin{eqnarray} \nonumber
  \tilde b_{\ell}(\eta) 
&=& \lim_{n\rightarrow \infty} \frac{1}{n} \log_2
\frac{ 
    [f(x)^{(1-R)n}]_{\ell n j}
    }
  {{n j \choose \ell n j}}
  {n \choose \ell n} \\ \nonumber
  &=& \lim_{n\rightarrow \infty} \frac{1}{n} \log_2  [f(x)^{(1-R)n}]_{\ell n j} \\ \label{asympt1}
  &-& \lim_{n\rightarrow \infty} \frac{1}{n} \log_2 {n j \choose \ell n j} 
  \hspace{-1mm}+\hspace{-1mm} \lim_{n\rightarrow \infty} \frac{1}{n} \log_2 {n \choose \ell n}.
\end{eqnarray}
By using the asymptotic approximation of a binomial
\begin{equation}
  \frac {1}{n}\log_2 {n \choose w} = H \left(\frac{w}{n} \right) + o(1),
\end{equation}
we have the following equality:
\begin{eqnarray} \nonumber
  - \frac{1}{n} \log_2 {n j \choose \ell n j} 
  \hspace{-1mm}+\hspace{-1mm} \frac{1}{n} \log_2 {n \choose \ell n}  
  \hspace{-3mm}
  &=&   \hspace{-3mm} - j H(\ell) + H(\ell) \hspace{-1mm}+\hspace{-1mm} o(1) \\
  &=&   \hspace{-3mm} - (j-1) H(\ell) \hspace{-1mm} +\hspace{-1mm} o(1),
\end{eqnarray}
where $o(1)$ is the term which converges to zero as $n \rightarrow \infty$.
Therefore, the sum of the second and the third terms in (\ref{asympt1}) converges to
$- (j-1) H(\ell)$.
The remaining task is evaluation of the first term in (\ref{asympt1}).
By using the Hayman method \cite{dru} \cite{modern} for evaluating the $n$-th coefficient of
a power of a polynomial, the following relation is obtained:
\[
\lim_{n \rightarrow \infty}\frac{1}{n} \log_2 [f(x)^{(1-R)n}]_{\ell j n}\hspace{20cm}
\]
\vspace{-10mm}
\begin{eqnarray} \nonumber
  &=& \lim_{n \rightarrow \infty}(1-R)\frac{1}{(1-R)n}  \log_2 [f(x)^{(1-R)n}]_{\ell k (1-R) n} \\ 
\nonumber
  &=& (1-R)\lim_{n \rightarrow \infty}\frac{1}{n} \log_2 [f(x)^{n}]_{\ell k n} \\ \nonumber
  &=&  (1-R)\lim_{n \rightarrow \infty} \frac{1}{n} \log_2 \frac{f(r)^{n}}{r^{\ell k n}} \\
  &=&  (1-R)(\log_2 (f(r)) - \ell k \log_2 r),
\end{eqnarray}
where $r$ is the 
smallest positive root of $x f'(x)/f(x) = \ell k$.
The claim of the lemma follows from the above results.
\hfill\qed
\end{lemma}
Since the asymptotic growth rate of other ensembles can be obtained 
with similar arguments, we omit the discussion of other ensembles.

In \cite{wadayama1}, a relation between the weight of a syndrome and 
the coset weight (defined in the next subsection)
is proved. The relation is called a {\em syndrome weight bound} in \cite{wadayama1}.
A similar bound can be derived for a regular bipartite ensemble.
\begin{lemma}[Syndrome weight bound for bipartite ensemble]
\label{synweight}
  If $\ell < \eta/k$ holds, then
  $\tilde b_{\ell}(\eta) = -\infty$. \\
(Proof) The polynomial $f(x)^{(1-R)n}$ can be expressed as
  \begin{equation}
    f(x)^{(1-R)n} = c_v x^v + c_{v+1} x^{v +1} + \cdots,
  \end{equation}
where $f(x) \defeq \alpha_k(x)^{1-\eta} \beta_k(x)^{\eta}$.
From the definition of $\alpha_k(x)$ and $\beta_k(x)$, we can see that
the minimum degree $v$ is equal to $\eta (1-R) n$.
It is obvious that $[f(x)^{(1-R)n}]_{\ell j n} = 0$ holds if $\ell j n < \eta (1-R) n$.
The above statement is equivalent to the claim of the lemma.
\hfill\qed
\end{lemma}

\begin{figure}[htbp]
\begin{center}
  \includegraphics[scale=0.7]{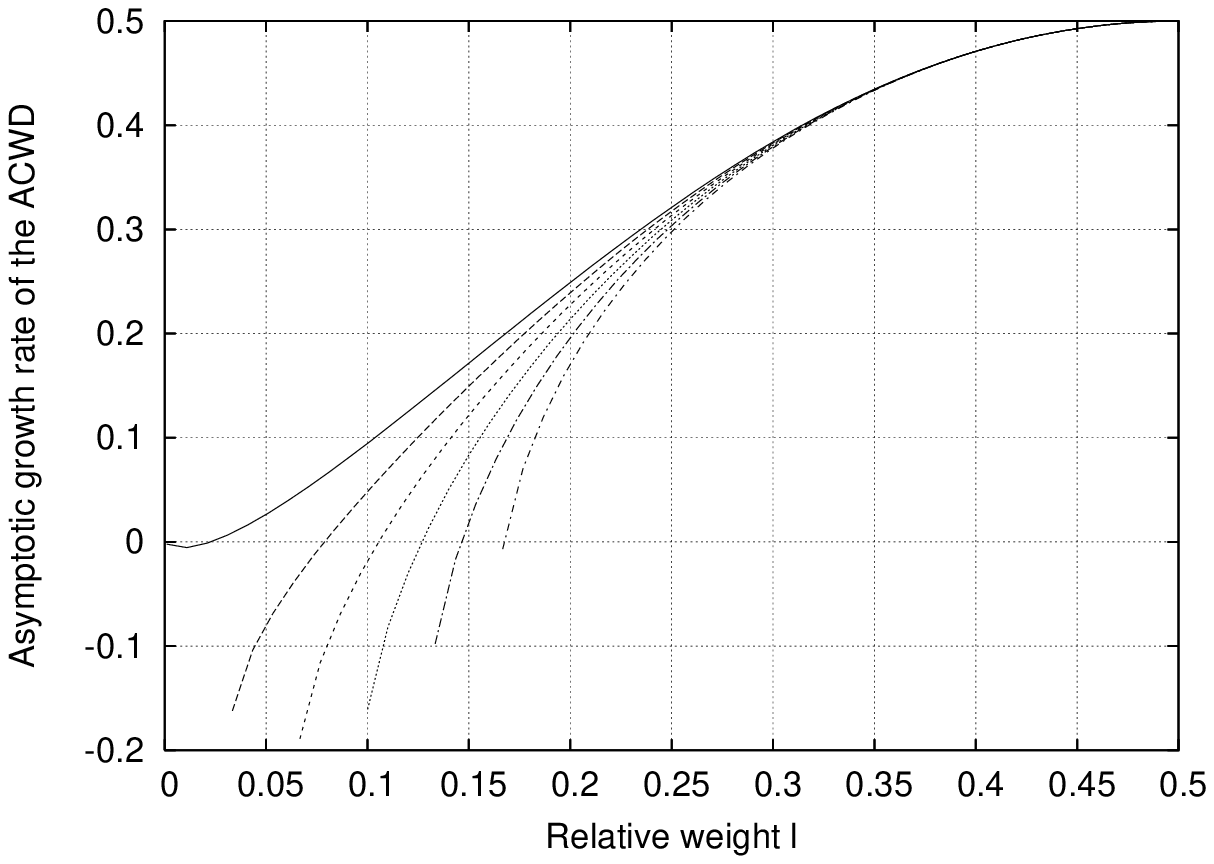} \\
{\small The curves in the above figure correspond to $\tilde b_{\ell}(\eta)$,
  $\eta = 0.0, 0.2, 0.4, 0.6, 0.8, 1.0$ from left to right.}
  \caption{AGR of $(j=3,k=6)$-regular bipartite ensemble}
  \label{fig:agr}
\end{center}
\end{figure}

\begin{example}
  Figure \ref{fig:agr} shows the AGR $\tilde b_{\ell}(\eta)$ of
  $(j=3,k=6)$-regular bipartite ensemble.
  For example,
  the rightmost curve corresponds to the case where $\eta = 1$. From 
  Lemma \ref{synweight}, we know $\tilde b_{\ell}(1) = -\infty$
  if $\ell < \eta/k = 1/6 \simeq 0.1667$. From Fig.\ref{fig:agr},
  we can observe that $\tilde b_{\ell}(1) > -\infty$ when
  $\ell \ge 1/6$.  \hfill\qed
\end{example}

\subsection{AGR of combined ensembles}

The next corollary reveals the relation between
the AGR of a concatenated ensemble and
those of component ensembles.
\begin{corollary}[AGR of concatenated ensembles]
  Consider two sequences of row symmetric ensembles  
  $X \defeq \{\Aen_n \}_{n>0}$ and $Y \defeq \{\Ben_n \}_{n>0}$.
  The ensemble $\Aen_n$ has column size $\nu_1 n (0 < \nu_1 < 1)$ 
  and row size $(1-R) n$ and
  $\Ben_n$ has column size $\nu_2 n (0 < \nu_2 < 1)$ 
  and row size $(1-R) n (0 < R < 1)$. We assume 
  that $\nu_1 + \nu_2 = 1$.

The asymptotic growth rate of the sequence of the concatenated ensembles
$Z \defeq \{\Aen_n \circ \Ben_n\}_{n>0}$  is given by
  \begin{equation}
    \tilde b_{\ell}^{(Z)} (\eta) 
    = \max_{(\ell_1, \ell_2, \kappa_1, \kappa_2) \in \Omega(\ell, \eta)}
    g(\ell_1, \ell_2,\kappa_1, \kappa_2),
  \end{equation}
  where $g( \ell_1, \ell_2, \kappa_1, \kappa_2)$ is defined by
  \begin{eqnarray} \nonumber
    g(\ell_1,\ell_2, \kappa_1, \kappa_2) 
    \hspace{-3mm}
    &\defeq&
    \hspace{-3mm}
    (1-R) \left(\eta H\left(\frac{\kappa_1}{\eta}\right) 
      \hspace{-1mm}+ \hspace{-1mm} (1-\eta) H\left(\frac{\kappa_2}{1-\eta}\right) \right) \\
    &+& \tilde b_{\ell_1}^{(X)}(\kappa_1 + \kappa_2) 
     + \tilde b_{\ell_2}^{(Y)}(\eta - \kappa_1 + \kappa_2)  
  \end{eqnarray}
  and $\Omega(\ell, \eta)$ is the set defined by
  \begin{eqnarray} \nonumber
    \Omega(\ell,\eta) \defeq
    \{(\ell_1,\ell_2, \kappa_1, \kappa_2): \ell = \nu_1 \ell_1 + \nu_2 \ell_2,0 \le \ell_1 \le 1, \\
    0 \le \ell_2 \le 1,0 \le \kappa_1 \le \eta, 0 \le \kappa_2 \le 1 - \eta \}.
  \end{eqnarray}
(Proof)
Let $m \defeq (1-R)n$.
From Corollary \ref{concatcoro}, the ACWD of the concatenated ensemble $\Aen_n \circ \Ben_n$ 
is given by
\begin{equation} \label{Qsum}
  \tilde B^{(\Aen_n \circ \Ben_n)}_{\ell n}(m \eta)
  =   \sum_{(\ell_1,\ell_2,\kappa_1,\kappa_2) \in \Omega} Q(\ell_1, \ell_2,\kappa_1, \kappa_2),
\end{equation}
where 
\begin{eqnarray} \nonumber
  Q(\ell_1, \ell_2,\kappa_1, \kappa_2) 
  \defeq {m \eta \choose m \kappa_1} {m(1-\eta) \choose m \kappa_2} \hspace{20mm} \\
  \times
  \tilde B^{(\Aen)}_{\ell_1 \nu_1 n}(m(\kappa_1+\kappa_2))  
  \tilde B^{(\Ben)}_{\ell_2 \nu_2 n}(m(\eta - \kappa_1+\kappa_2)).
\end{eqnarray}
The sum in (\ref{Qsum}) can be upper bounded by the product of 
the number of terms and the maximum value of the summands. Since the number of terms are 
bounded by $n^3$, we have the following upper bound:
\begin{equation} \label{upperbound}
  \tilde B^{(\Aen_n \circ \Ben_n)}_{\ell n}(m \eta)
  \le n^3 \max_{(\ell_1,\ell_2, \kappa_1,\kappa_2) \in \Omega } Q(\ell_1, \ell_2,\kappa_1, \kappa_2).
\end{equation}
On the other hand, the sum in (\ref{Qsum}) can be lower bounded by any single term. Namely, we have
\begin{equation} \label{lowerbound}
  \tilde B^{(\Aen_n \circ \Ben_n)}_{\ell n}(\eta m)  
  \ge \max_{(\ell_1,\ell_2, \kappa_1,\kappa_2) \in \Omega } Q(\ell_1, \ell_2,\kappa_1, \kappa_2).
\end{equation}
From the upper bound (\ref{upperbound}) and the lower bound (\ref{lowerbound}), we immediately
obtain
\begin{eqnarray}  \nonumber
  \frac 1 n \log_2 
  \tilde B^{(\Aen_n \circ \Ben_n)}_{\ell n}(\eta m)
  \hspace{-3mm}
  &=& 
  \hspace{-3mm}
  \frac 1 n \log_2 
  \max_{(\ell_1,\ell_2, \kappa_1,\kappa_2) \in \Omega } Q(\ell_1, \ell_2,\kappa_1, \kappa_2) \\
  &+&   \hspace{-3mm} o(1).
\end{eqnarray}
The claim of the corollary is obtained by
\begin{eqnarray}  \nonumber
  \tilde b_{\ell}^{(Z)}(\eta)  
  \hspace{-3mm}
  &=&
  \hspace{-3mm}
  \lim_{n \rightarrow \infty} \frac 1 n \log_2 
  \tilde B^{(\Aen_n \circ \Ben_n)}_{\ell n}(\eta m) \\ \nonumber
  &=&
  \hspace{-3mm}
  \lim_{n \rightarrow \infty} \left(\frac 1 n \log_2 
    \max_{(\ell_1,\ell_2, \kappa_1,\kappa_2) \in \Omega } 
    Q(\ell_1, \ell_2,\kappa_1, \kappa_2) + o(1) \right)\\ 
     &=&    \hspace{-6mm} \max_{(\ell_1,\ell_2, \kappa_1,\kappa_2) \in \Omega } 
     g(\ell_1, \ell_2,\kappa_1, \kappa_2).
  \end{eqnarray}
  \hfill\qed
\end{corollary}
\begin{remark}
  Based on similar arguments, we can prove the asymptotic counterparts of
  Corollary \ref{stackacwd} for stacked ensembles, Corollary \ref{typeI} for type I 
  combined ensembles, and Corollary \ref{col_type2} for type II combined ensembles. 
  The essential part of such proofs is that
  the number of terms corresponding to a summation in an ACWD formula 
  is polynomial order to $n$ (e.g., Corollary \ref{typeI}).
  Thus, we can replace a summation operator by a max operator 
  to obtain the asymptotic growth rate. \hfill\qed
\end{remark}

\subsection{Typical coset weight}

From the ACWD of an ensemble, we can obtain detailed information
on an instance of the ensemble. We here discuss the weight of
the coset leader, which is called {\em coset weight}.
The following discussion is based on the standard argument on 
the typical minimum distance \cite{Gal63}. 

For a given parity check matrix $H$ and a syndrome $\ve s$, 
the coset weight $D(H, \ve s)$ is defined by
\begin{equation}
  D(H, \ve s) \defeq \min_{\ve x \in C(H,\ve s)} |\ve x|.
\end{equation}
The accumulated coset weight distribution is defined by 
\begin{equation}
  F_\tau(H, \ve s) \defeq \sum_{w = 0}^\tau A_w(\ve s, H),\quad \tau \in [0,n].
\end{equation}
From the above definitions, it is clear that $D(H,\ve s) < \tau$
if $F_\tau(H,\ve s) \ge 1$ holds.

Assume that a syndrome $\ve s$ is given and
we will draw a parity check matrix from an ensemble according to 
the assigned probabilities.
By using the Markov inequality, it can be proved that
\begin{equation}
  \label{markov}
  Pr[F_\tau(H,\ve s) \ge 1] \le \tilde F_\tau (\ve s),
\end{equation}
where the average of the accumulated coset weight distribution $\tilde F_\tau (\ve s)$ 
is given by
\begin{eqnarray} \nonumber
  \tilde F_\tau (\ve s) &=& \sum_{H \in \Gen} P(H) F_\tau (H,\ve s) \\ \nonumber
%  &=& \sum_{H \in \Gen} P(H) \sum_{w = 0}^\tau A_w(\ve s, H) \\ \nonumber
  &=& \sum_{w = 0}^\tau \sum_{H \in \Gen} P(H)  A_w(\ve s, H) \\
  &=& \sum_{w = 0}^\tau \tilde A_w (\ve s).
\end{eqnarray}
The equality (\ref{markov}) means that 
the probability such that a coset with 
coset weight smaller than $\tau$ is sampled is upper bounded by $\tilde F_\tau (\ve s)$.
Namely, evaluation of $\tilde F_\tau (\ve s)$ brings us information on the coset weight
$D(H,\ve s)$.

For analyzing the asymptotic behavior of $\tilde F_\tau (\ve s)$, we introduce 
the {\em typical coset weight} in the following way:
\begin{definition}[Typical coset weight]
  Let $\tilde b_\ell(\eta) (0 \le \ell \le 1, 0 \le \eta \le 1)$ be the AGR
  of a given ensemble. The typical coset weight of the ensemble is defined by
\begin{equation} \label{inf}
  \theta_{\eta'} \defeq \arg \min \{\ell: \tilde b_\ell(\eta') \ge  0, 0 \le \ell \le 1 \}
\end{equation}
for given $\eta' (0 \le \eta' \le 1)$. \hfill\qed
\end{definition}
For example, in the case of a regular bipartite ensemble, (\ref{inf}) is equivalent to 
$
  \theta_{\eta'} = \arg \min \{\ell: \tilde b_\ell(\eta') =  0, 0 \le \ell \le 1 \},
$
which is the smallest root of $\tilde b_\ell(\eta') =  0$.

Since we will only deal with a row symmetric ensemble in the following,
we define $G_\tau(|\ve s|,H) \defeq \tilde F_\tau (\ve s,H)$ and
$G_\tau(|\ve s|) \defeq \tilde F_\tau (\ve s)$.

The following theorem shows that 
the probability such that a coset with coset weight smaller than $\ell'n (\ell' < \theta_\eta)$ 
is sampled converges to zero as $n$ goes to infinity.
\begin{theorem}
Suppose that $\ell' \defeq \theta_\eta - \epsilon$, where $\epsilon$ is any positive 
real number satisfying $\ell' \ge 0$. 
The equality 
\begin{equation}
  \lim_{n \rightarrow \infty} Pr[\tilde G_{\ell' n}(H, \eta (1-R)n) \ge 1] = 0
\end{equation}
holds for given $\eta (0 \le \eta \le 1)$. \\
(Proof)
From this definition of $\theta_\eta$, we can see that 
$\tilde b_{l}(\eta) < 0$ holds for $0 \le l \le \ell'$.

In order to evaluate the asymptotic value of $\tilde G_{\ell' n}(\eta (1-R)n)$, 
we consider $(1/n) \log_2 \tilde G_{\ell' n}(\eta (1-R)n)$, which is simplified
in such a way:
\[
\frac{1}{n} \log_2 \tilde G_{\ell' n}(\eta (1-R)n) \hspace{40mm}
\]
\vspace{-6mm}
\begin{eqnarray} \nonumber
  &=& 
  \frac{1}{n} \log_2  \left(\sum_{w=0}^{\ell' n} \tilde B_w(\eta (1-R)n)\right) \\ \nonumber
  &\le&
  \frac{1}{n} \log_2  \left((\ell' n +1 ) \max_{w=0}^{\ell' n} 
    \tilde B_w(\eta (1-R)n)\right)  \\ \nonumber
  &=&  
  \max_{l=0}^{\ell' } \frac{1}{n} \log_2  \tilde B_{l n}(\eta (1-R)n) + o(1).
\end{eqnarray}
From the above inequality and the inequality  $\tilde b_{l}(\eta) < 0 (0 \le l \le \ell')$,
we  have 
\begin{eqnarray} \nonumber
  \lim_{n \rightarrow \infty}\frac{1}{n} \log_2 \tilde G_{\ell' n}(\eta (1-R)n) 
  &\le&  \max_{l=0}^{\ell' }  \tilde b_{l}(\eta) \\
  &<& 0.
\end{eqnarray}
This implies 
\begin{equation}
  \lim_{n \rightarrow \infty} \tilde G_{\ell' n}(\eta (1-R)n) = 0.
\end{equation}
The use of the Markov inequality (\ref{markov}) and this equality leads to 
the claim of the theorem.
\hfill\qed  
\end{theorem}

\begin{example}
  In the case of (3,6)-regular bipartite ensemble, we have
  $\theta_{0.2} \simeq 0.0788$ and $\theta_{0.8} \simeq 0.146$. In general,
  it is observed that the typical coset weight increases as $\eta$ grows (see also Fig.\ref{fig:agr}).
  \hfill\qed
\end{example}

\section*{Acknowledgment}
The author would like to thank Mr. Kenta Kasai for inspiring discussion and
comments to \cite{wadayama0}. The author also thank Mr. Ryoji Ikegaya for 
their preprint \cite{Ikegaya}.

\section*{Appendix}

(Proof of Lemma \ref{bipartite}) Consider the socket model\cite{MV04} for $(j,k)$-regular bipartite 
graph ensemble. The symbols $j$ and $k$ denote the variable
node degree and the check node degree, respectively.
The check node sockets are denoted by $(c_1,c_2,\ldots,c_{jn})$ and the  
variable node sockets are denoted by $(v_1,v_2,\ldots,v_{jn})$.
The bipartite graph ensemble consists of all the bipartite graphs 
constructed in the following way:
Let $\pi_{jn}$ be the set of all permutation on $jn$ elements.
$
  (u_1,u_2,\ldots u_{jn}) = \pi((v_1,v_2,\ldots,v_{jn}))
$
A bipartite graph $G(\pi) (\pi \in \pi_{jn})$ is obtained
by connecting $c_i$ and $u_i (i \in [1,jn])$ with an edge. 
There is one-to-one correspondence between a permutation in $\pi_{jn}$
and a $(j,k)$ regular bipartite graph. Thus the size of 
$\Gen$ is equal to $(jn)!$.

Assume that $w$ variable nodes corresponds to 1 as their value.
The number of edges connecting such variable nodes (called active
edges) is equal to $jw$. The rest of edges are called inactive edges
and the number of inactive edges is $jn-jw$.

The edge configuration is a binary vector $(e_1,e_2,\ldots,e_{jn})$
where $e_i = 1$ if the edge from $c_i$ is active;
$e_i = 0$ otherwise.
The number of possible edge configuration $E(\ve s)$ which 
corresponds to a syndrome $\ve s$ is
given by
\begin{equation}
  E(\ve s) = 
  \left[\alpha_k^{m-|\ve s|}(x) \beta_k^{|\ve s|}(x)\right]_{w j}.
\end{equation}
For a given edge configuration,
there are $(jn - jw)! (jw)!$ graphs which yields the same configuration.
This is because the permutations within the 
active edges and inactive edges do not change the 
edge configuration. We now use the one-to-one correspondence 
between a graph and matrix. This correspondence leads to 
the claim of the lemma in the following way:
\begin{eqnarray} \nonumber
  \tilde A_w(\ve s) 
  &=& \frac{\#\{H\in \Gen: H \ve z^t = \ve s \}}{\# \Gen} {n \choose w} \\ \nonumber
  &=& \frac{\left[\alpha_k^{m-|\ve s|}(x) 
      \beta_k^{|\ve s|}(x)\right]_{w j} (jn - jw)! (jw)!}{(jn)!} 
  {n \choose w} \\
  &=&   \frac{\left[
        \alpha_k^{m-|\ve s|}(x) \beta_k^{|\ve s|}(x)
      \right]_{w j}}{{n j \choose w j}} {n \choose w}. 
\end{eqnarray}
\hfill\qed

\end{document}